\documentclass[11pt]{article}
\usepackage{pifont}
\usepackage{natbib}
\usepackage{graphicx}
\usepackage{amssymb}
\usepackage{epstopdf}
\usepackage{amsmath}
\usepackage{lineno}
\usepackage{setspace}
\usepackage{url}
\usepackage{lscape}
\usepackage{color}
\usepackage[normalem]{ulem}
\usepackage{sidecap}
\usepackage[font={small}]{caption}
\usepackage{authblk}
\usepackage{fancyhdr}

\begin{document}

\title{Tsunami Bores in Kitakami River}
\author[1]{Elena Tolkova \thanks{e.tolkova@gmail.com; elena@nwra.com}}
\author[2]{Hitoshi Tanaka \thanks{hitoshi.tanaka.b7@tohoku.ac.jp}}
\affil[1]{NorthWest Research Associates, Bellevue, WA 98009-3027, USA}
\affil[2]{Department of Civil Engineering, Tohoku University, 6-6-06 Aoba, Sendai, 980-8579, Japan}
\date{}
\maketitle

\pagestyle{fancy}

The final publication is available at Springer via {\url{http://dx.doi.org/10.1007/s00024-016-1351-7}}\\

Cite this article as:
Tolkova, E. \& Tanaka, H. Pure Appl. Geophys. (2016) 173: 4039. https://doi.org/10.1007/s00024-016-1351-7

\tableofcontents

\begin{abstract}
The 2011 Tohoku tsunami entered the Kitakami river and propagated there as a train of shock waves, recorded with a 1-min interval at water level stations at Fukuchi, Iino, and the weir 17.2 km from the mouth, where the bulk of the wave was reflected back.  
The records showed that each bore kept its shape and identity as it traveled a 10.9-km-path Fukuchi-Iino-weir-Iino. 
Shock handling based on the cross-river integrated classical shock conditions was applied to reconstruct the flow velocity time histories at the measurement sites, to estimate inflow into the river at each site, to evaluate the wave heights of incident and reflected tsunami bores near the weir, and to estimate propagation speed of the individual bores. Theoretical predictions are verified against the measurements.
We discuss experiences of exercising the shock conditions with actual tsunami measurements in the Kitakami river, and test applicability of the shallow-water approximation for describing tsunami bores with heights ranging from 0.3 m to 4 m in a river segment with a depth of 3-4 m. 
\end{abstract}

\section{Introduction}

Rivers are known to be the ``tsunami highways". Tsunamis penetrate in rivers much farther inland than the coastal inundation reaches over the ground, and can cause flooding in low-lying areas located several km away from the coastline. The evidence of tsunami penetration in river has been described by Abe (1986) and Tsuji et al. (1991) after the 1983 Japan Sea tsunami, and Yasuda (2010) after the 2003 Tokachi-Oki tsunami.  Tanaka et al (2008) analyzed damages caused by the 2004 Indian Ocean tsunami in five rivers in Sri Lanka, where the tsunami intruded from 3-4 to 20 km upstream and damaged 34 pedestrian, road, and railway bridges altogether. The recent 2010 Chilean and 2011 Tohoku  trans-Pacific tsunamis penetrated into rivers around the ocean. The 2010 Chilean tsunami caused by a Mw 8.8 earthquake was 10-30 m high by the coast of Chile, where it propagated at least 15 rkm (river-km, distance along the river from its mouth) up the Maule river. Offshore fishing boats were carried upriver and deposited on the river banks for as far as 10 rkm  \citep{fritz2011}. After crossing the Pacific, the tsunami came to the northeast coast of Japan about 1 m high,  penetrated a few rivers, and was recorded, in particular, at a water level station in Old Kitakami river at a 22 rkm point  \citep{kayane, tanaka2014}. 
The longest runups in the 2011 Tohoku tsunami after Mw 9.0 earthquake occurred along rivers, where inundation by rivers reached through 3-4 times greater distances, than inundation spreading over the ground in the same area  \citep{morisurvey, tanaka2014, liu2013}. Across the Pacific, this tsunami was detected in the Columbia river 100+ km up the river mouth \cite[]{yeh}. Unique features of tsunami dynamics in rivers were brought together and described by Tolkova et al. (2015).

Photo and video evidence sometimes shows tsunamis propagating in rivers as bores or undular bores.  
This appealing feature, however, had not been captured in the field measurements described in the literature.   
Bores are shock waves with nearly vertical fronts separating flows of different depths, which are sometimes followed by a train of short (of order of tens of meters, or with periods of a few seconds) waves, or undulations. Resolving a temporal structure of a passing bore requires an instrument with a high-frequency sampling rate.  
Tsunami measurements are commonly provided by water level stations, such as tide gauges. In Japan, water levels in main rivers are monitored by the Ministry of Land, Infrastructure, Transportation, and Tourism (MLIT) with a network of stations reporting with a 10 min interval, and only in recent years some stations started reporting at a 1 min interval. During the 2011 tsunami, these stations with a 1-min sampling rate recorded the tsunami passage along the Kitakami river. The records clearly show a train of shock waves, with each wave being trackable through three locations on its journey upriver to a weir at 17 rkm and back. 
In this work, we analyze the tsunami bore kinematics as inferred from the measurements, and compare with predictions based on the classical (hydrostatic) shock theory, thus testing applicability of the shallow-water approximation for describing tsunami bores with heights ranging from 0.3 m to 4 m in a river segment with a depth of 3-4 m. 
Shock handling based on the cross-river integrated classical shock conditions was applied to reconstruct the flow velocity time histories at the measurement sites, to estimate inflow into the river at each site, to evaluate the wave heights of incident and reflected tsunami bores near the weir, and to estimate propagation speed of the individual bores. As we proceed, we discuss experiences of exercising the shock conditions with actual measurements of a tsunami in a real river.

\section{Discussion of Observations}

\begin{figure}[ht]
	\resizebox{\textwidth}{!}
		{\includegraphics{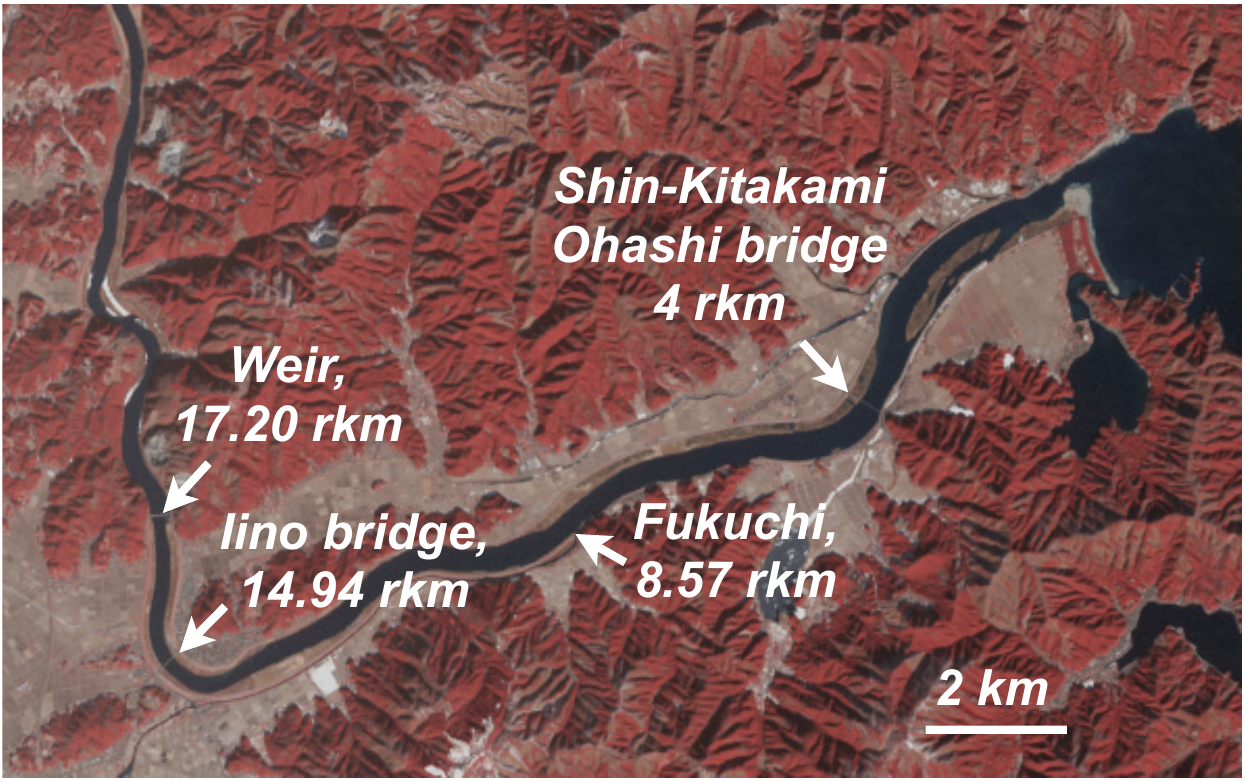}}
	\caption{ Arial view of lower Kitakami river on January 16, 2011, with locations of observation points. Image credit to NASA's Earth Observatory.}
	\label{map}
\end{figure}

\begin{figure}[ht]
	\resizebox{\textwidth}{!}
		{\includegraphics{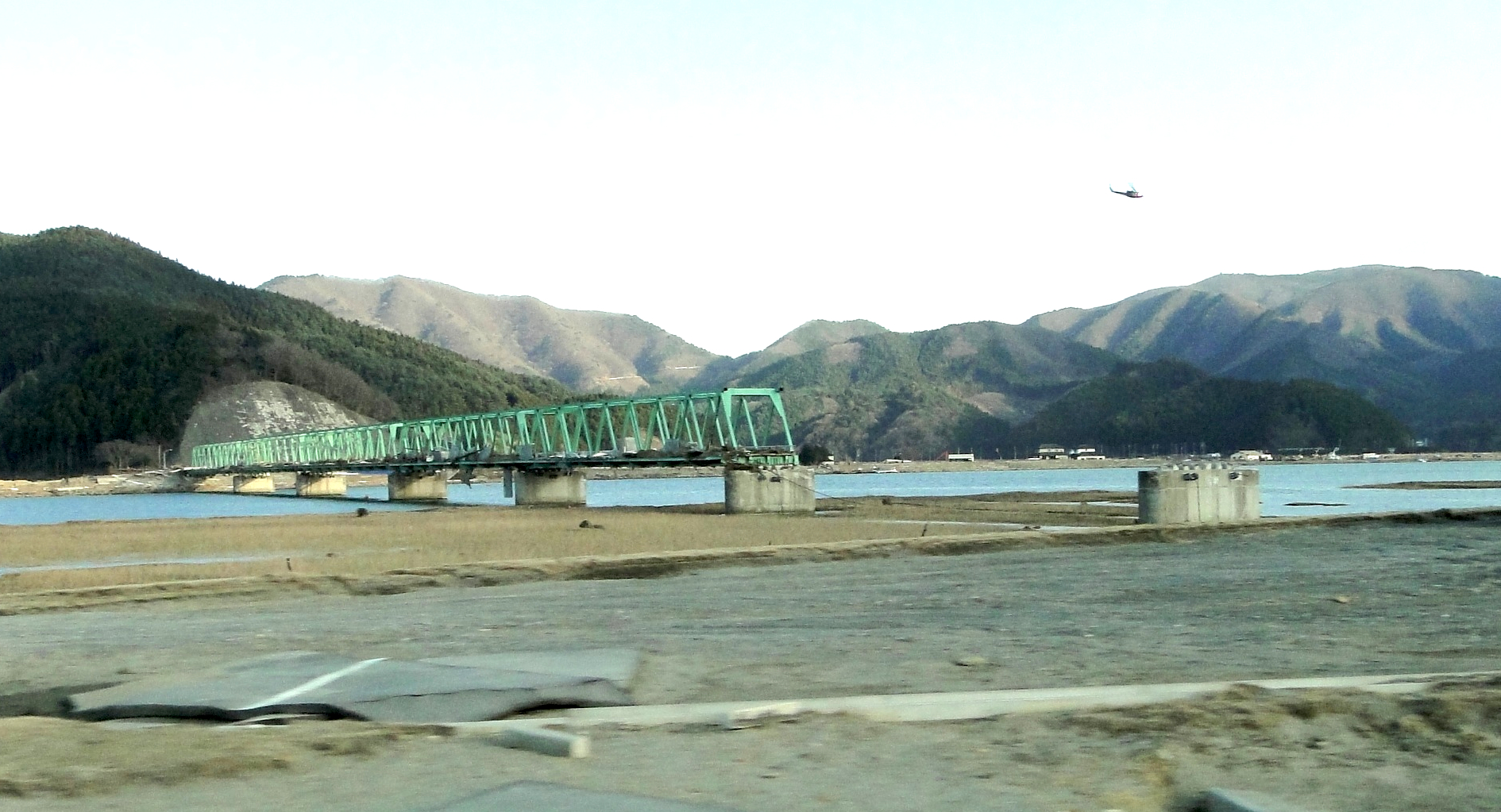}}
	\caption{Shin-Kitakami Ohashi Bridge across the Kitakami river at 4 km from the mouth with a washed out section, after the 2011 Tohoku tsunami. Photo by H. Tanaka.}
	\label{bridge}
\end{figure}
\begin{figure}[ht]
	\resizebox{\textwidth}{!}
		{\includegraphics{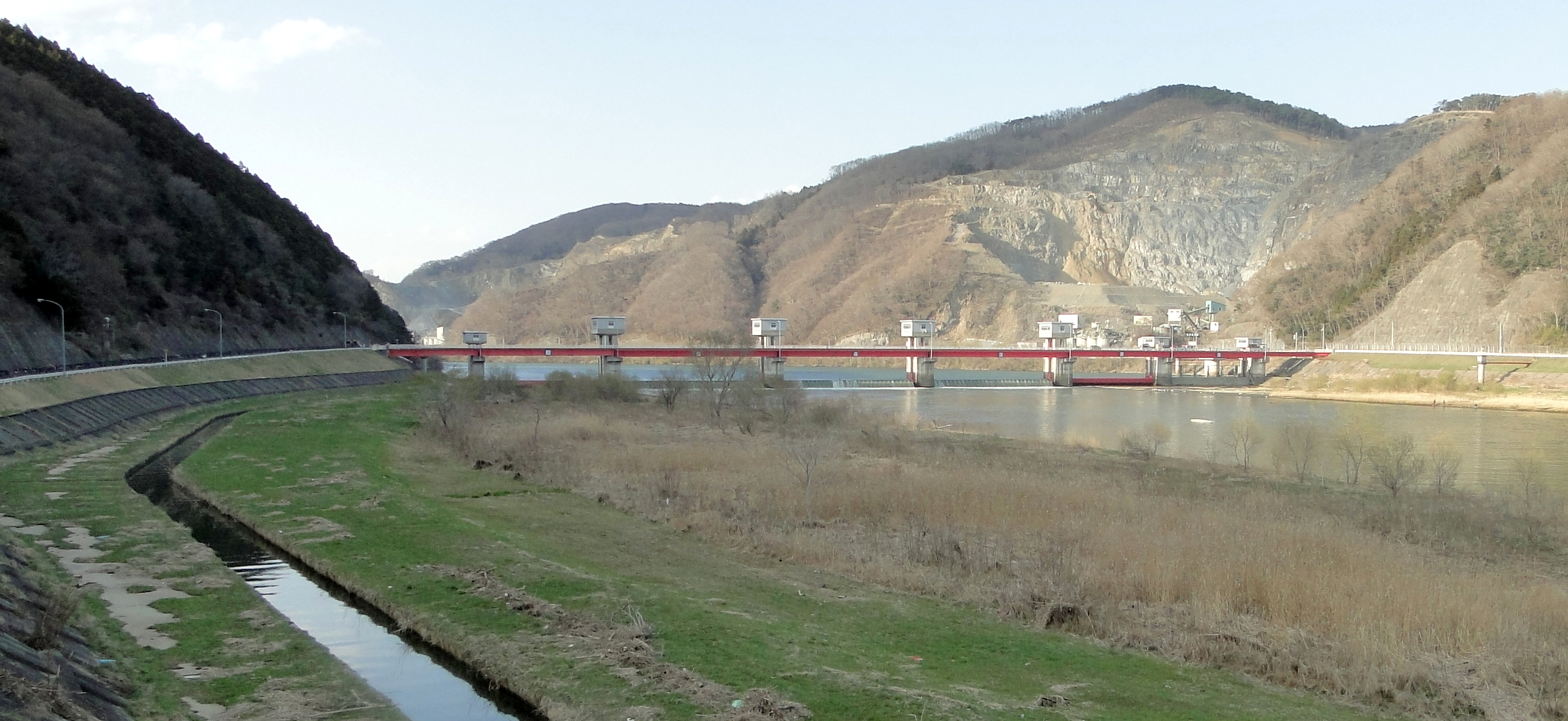}}
	\caption{Weir in the Kitakami River. Photo by H. Tanaka.}
	\label{weir}
\end{figure} 

\begin{figure}[ht]
	\resizebox{1.1\textwidth}{!}
		{\includegraphics{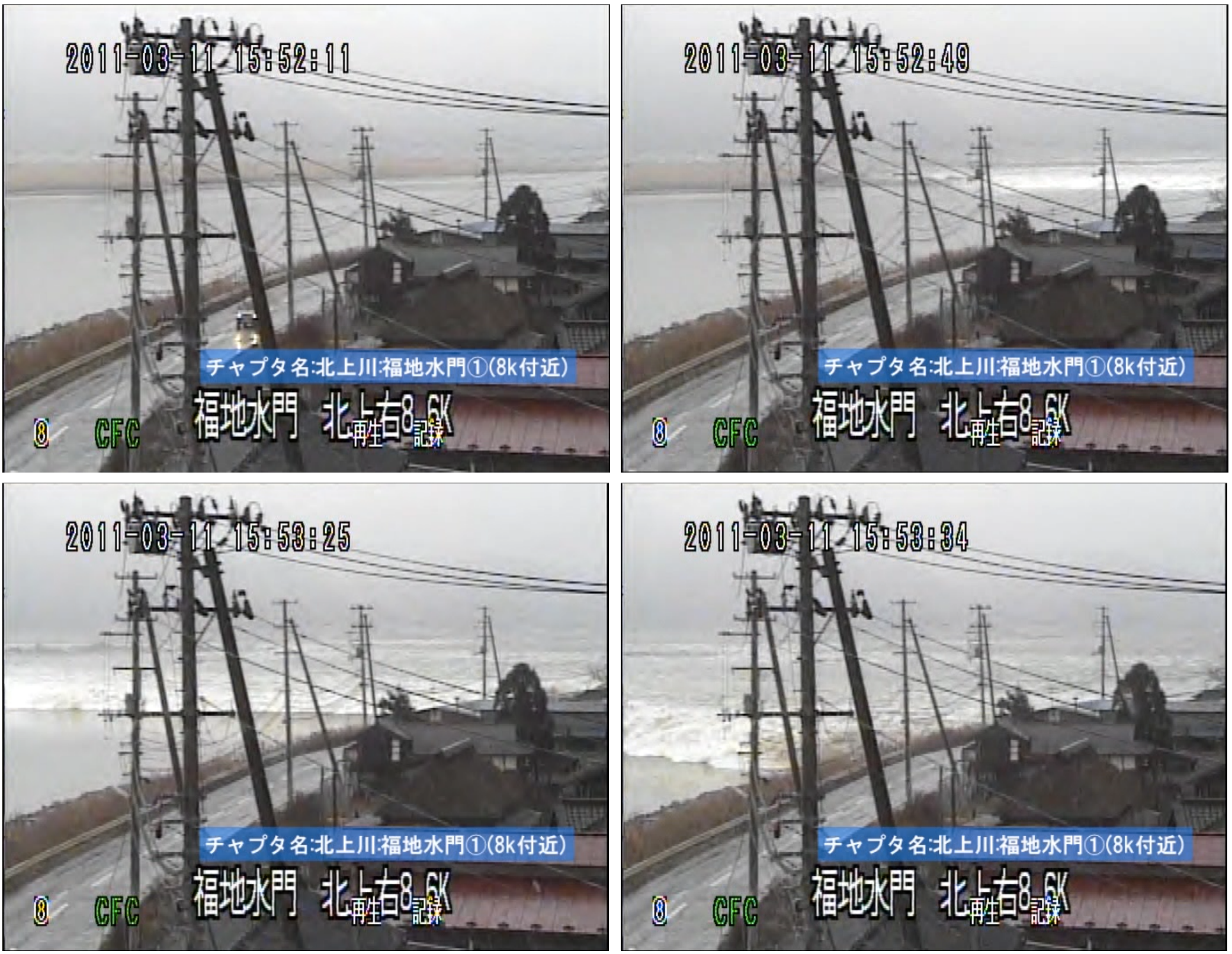}}
	\caption{ Tsunami passing Fukuchi, on a surveillance video record (the clock on the images is offset). Left (north) shore disappears under water, as the tsunami passes by. In spite of highly non-uniform flow depth, the tsunami front tends to remain straight across the river.  Time intervals between the frames are 38, 36, and 9 sec.}
	\label{video}
\end{figure}

The longest tsunami intrusion distance in Japan in the 2011 was observed in the Kitakami river (Figure \ref{map}) - the fourth largest river in Japan and the largest on the northeast coast of Honshu. 
The tsunami  washed out a section of a steel 6.5-m high bridge at 4 km from the river mouth (Figure \ref{bridge}), and inundated lands for 6 km along the river. Past 6 rkm, the tsunami was mostly confined to the river and its flood channels bordered by 5+ m high dikes. In about 30 min after entering the river, tsunami came to a movable weir at a 17 rkm mark (Figure \ref{weir}) built to prevent salt damage to domestic, industrial and irrigation water supply lines, as  well as to regulate the river flow. The weir stayed 3.6 m high above the river surface on the downstream side, and the gates were closed for the tsunami. 
The bulk of the tsunami energy was reflected back, but a few waves overtopped the weir and, greatly reduced in height, continued upriver and reached up to 49 rkm, where the river bed elevation was 4.6 m above the sea level \cite[]{tanaka2014}.

The tsunami passage up Kitakami can be tracked on water level time histories registered at Fukuchi at 8.57 rkm, Iino at 14.94 rkm, and the weir at 17.20 rkm. The location of the measurement stations along the river is shown in Figure \ref{map}. 
The tsunami records with 1-min sampling interval show that from Fukuchi to the weir, the tsunami traveled as a train of shock waves (bores), also seen in a few video records 
(Figure \ref{video}). On this journey, the bores were remarkably steady and kept their identity, so they can be individually traced through the records from Fikuchi to Iino to the weir, and then back to Iino after reflecting from the weir. Most of the reflected bores entirely dissipated before they could return to Fukuchi. 

\begin{figure}[ht]
	\resizebox{\textwidth}{!}
		{\includegraphics{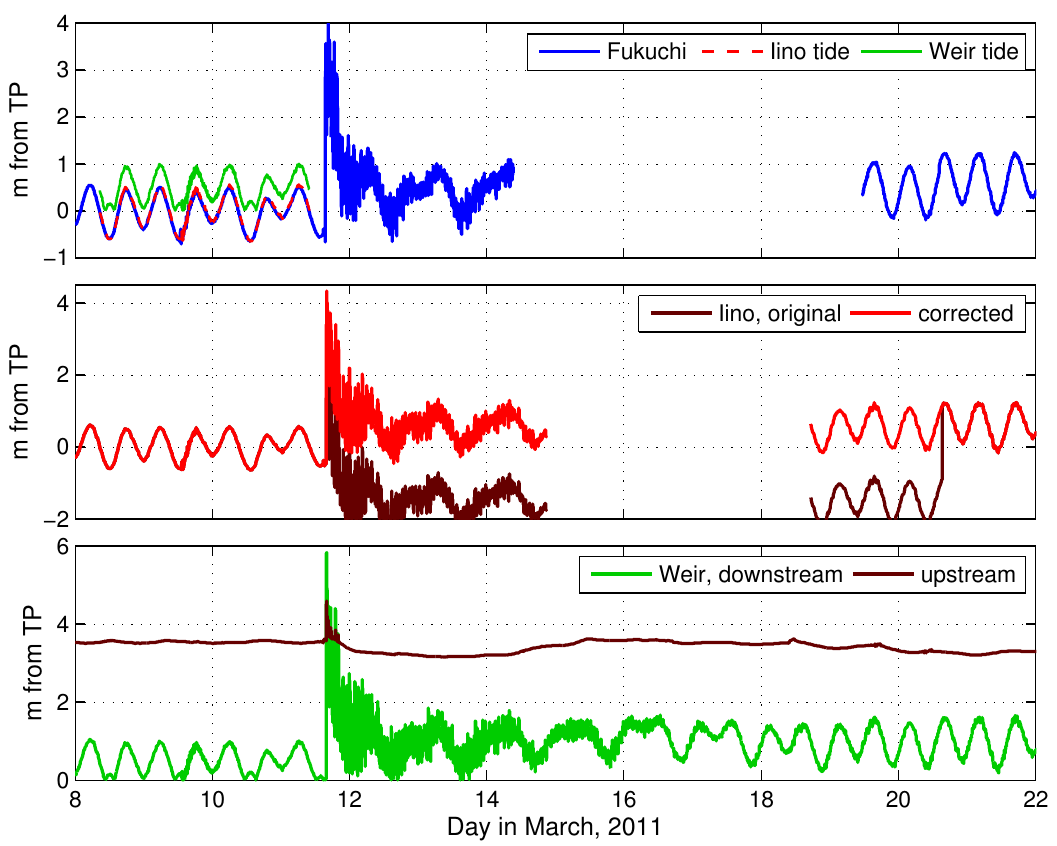}}
	\caption{Water level time histories at Fukuchi, Iino, and downstream and upstream the weir. For deducing the tidal regime, Fukuchi record is overlaid with a 3-day-long segments of tidal records by Iino and by the weir. The original Iino record had a vertical shift in the middle segment, starting with the tsunami arrival. }
	\label{recs}
\end{figure}

\begin{figure}[ht]
	\includegraphics[width=\textwidth]{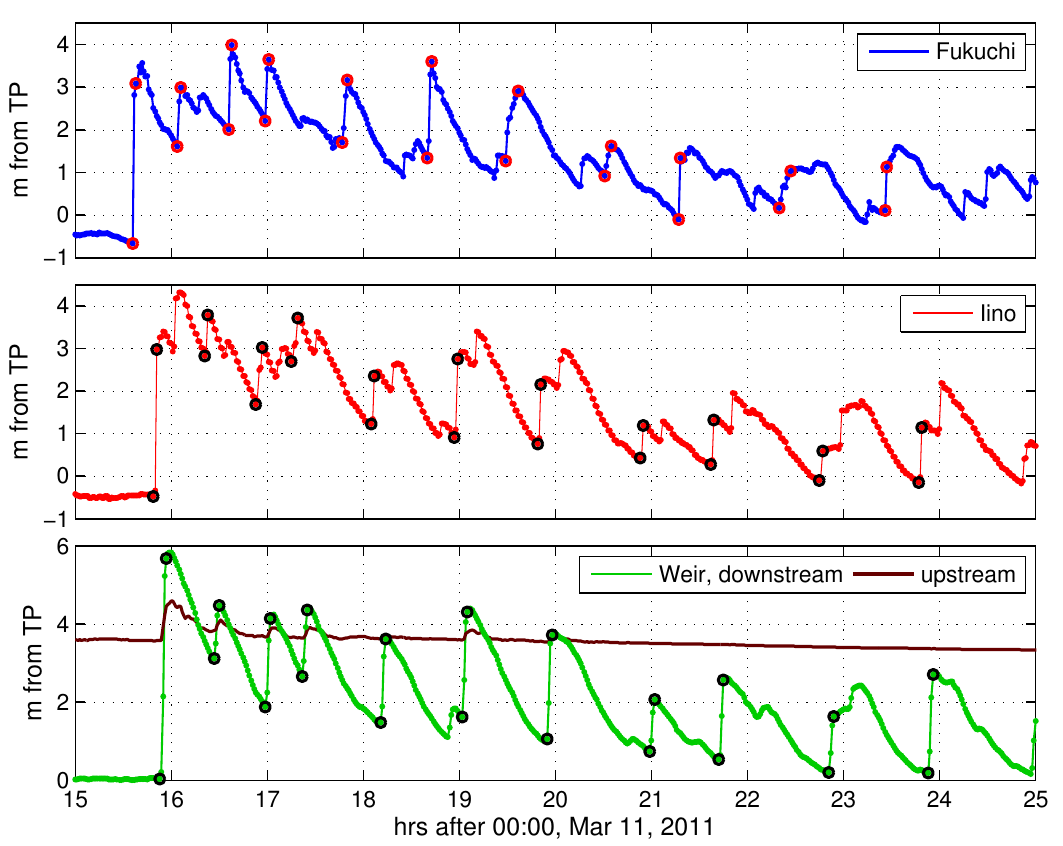} \\
	\includegraphics[width=\textwidth]{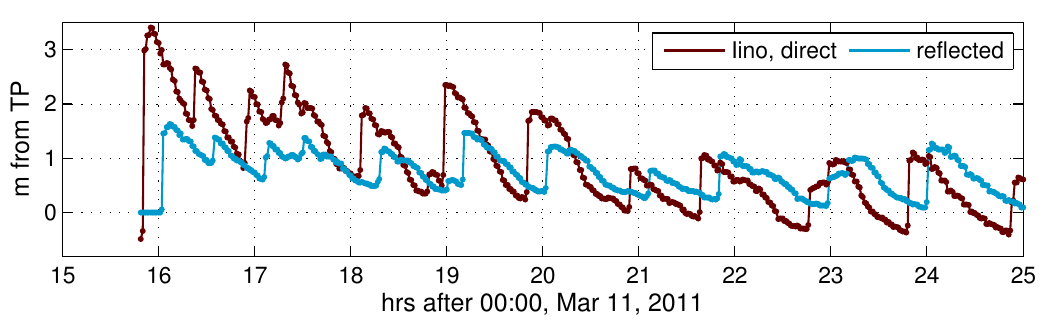} 
	\caption{Top three plots: zoomed-in records showing the first eleven tsunami waves at the locations. Circles (red or black) mark readings at the foot and on the top of each upriver-going shock. Bottom: Iino record decomposed into direct and reflected wave-trains.  }
	\label{zoom}
\end{figure}

The tsunami records are shown in Figure \ref{recs}, zoomed-in in Figure \ref{zoom}, with elevations relative to TP, and the time being the Japanese Standard Time. The first and tallest tsunami bore passed Fukuchi 4.5-m high, Iino 3.4-m high, and created a 6-m high wave by the weir, 2.2 m above its top. A 1-m-high wave continued upriver. Tsunami force displaced the gate and caused its leakage, as seen in variations of the upstream water level after the event. Tsunami also caused an offset of Iino gage readings (dark red in Figure \ref{recs}, middle plot). It appears that applying a uniform correction to the shifted 9-day-long data segment repairs the problem (red in Figure \ref{recs}, middle plot). Should, however, the offset not be uniform, but contain an initial trend, then the height of the first bore by Iino would be determined with an error. 
It is also observed, that the tsunami set a new mean river stage (elevated by 0.6 m), dictated by the new mean sea level relative to the land subsided in the earthquake \cite[]{tanakayeh}. There is also an apparent super-elevation of the short-term mean river stage during the main tsunami activity manifesting a so-called backwater effect \cite[]{tolkova2015}, also observed in relation to tides intruding into rivers \citep{buschman2009, buschman2010}. We also note, that tidal variations at the stations, when not obscured by tsunami, occur in co-phase; in particular, tidal signals at Fukuchi and Iino are visually identical (Figure \ref{recs}, top). Hence tide forms a standing wave in the river, owing to the presence of the weir. Tidal current is zero at low or high tide, and maximal in-between.
The low tide took place at 13:30, 2 h before the tsunami arrival. Assuming a 12-h tidal cycle, maximal tidal inflow current occurred around 16:30, ebb began at 19:30, and tidal outflow current was maximal at 22:30.

The first eleven tsunami waves, which are the focus of this study, are displayed in zoomed-in records in Figure \ref{zoom}. 
There are four measurement stations at the weir, which are located by the right and left banks, on the upstream and downstream side. The records on the left and right sides downstream the weir are visually identical, which suggests that the waveforms were uniform across the river, in spite of significant cross-river variations of the flow depth.  
All waves have nearly vertical fronts (shocks), marked with circles around readings delimiting each upriver-propagating shock. The wave-train forms are remarkably alike at Fukuchi and in front of the weir. The waveform in the Iino record looks different because of the overlap of the direct wave train and that reflected from the weir. The Iino record was decomposed into direct and reflected wave-trains, as described in the Appendix. The resulting direct and reflected component in Iino record shown in Figure \ref{zoom}, bottom plot, each has again the same waveform as recorded at the other two locations. We conclude, therefore, that each bore kept its shape and identity as it traveled 10.9 km between Fukuchi-Iino-weir-Iino. 

\clearpage

\section{Bore propagation speed}
\label{pspeed}

An average wave celerity between Iino and the weir for each shock, direct or reflected, was computed given each shock's travel time between Iino and the weir ($2,260$ m), in a respective direction. The travel times were determined by the arrival times of shock fronts at the locations, as marked in Figure \ref{zoom} by the lower point in each pair.
The resulting upriver bore celerities at Iino are found to be 9.4,  6.3,  6.3,  5.4,    6.3, 7.5,    6.3,    6.3,    7.5,    6.3, and 6.3 $m/s$; while after reflection, these bores traveled back at  4.7,  6.3, 6.3, 5.4, 6.3, 5.4,    6.3,    6.3,    5.4,    6.3, and 6.3 $m/s$. These variations of the propagation speed can be attributed to the flow depth variations, as well as to slowing a particular bore down (with respect to the shores) by the current of another bore propagating in the opposite direction. 
On the contrary, it took the same 12 min for each bore to travel from Iino to the weir and back to Iino, with the corresponding average celerity being 6.3 $m/s$.
In that case, the contribution of the background flow into the average wave celerity on the way Iino-weir-Iino was largely nulled due to back and forth passage.

The eleven bores took, respectively, 13, 17, 17, 16, 18, 17,  20,  22,  20,  25, and 21 min to travel $6,370$ m between Fukuchi and Iino, which corresponds to average Fukuchi-Iino propagation speeds of 8.2, 6.2, 6.2, 6.6, 5.9, 6.2, 5.3, 4.8, 5.3, 4.2, and 5.1 $m/s$.
Variations of the tsunami propagation speed relative to the shore might correlate with the variations of the background flow.  
In particular, Fukuchi-Iino travel times went up for the waves passing Fukuchi on or after 19:30 with the beginning of tidal ebb and drawdown of the accumulated tsunami-brought water. The slowest 10-th tsunami wave travelled with the maximal outgoing tidal current around 22:30.
Also, all direct waves between Fukuchi and Iino, except the first one, might have been slowed down by the opposing current in the reflected waves. For future analysis, each bore celerity at Iino will be approximated by this bore's average celerity between Iino and the weir, and celerity at Fukuchi -- by the bore's average celerity between Fukuchi and Iino.

\section{Shock equations in a river}
\label{algebra}

\begin{figure}[ht]
\centering
	\resizebox{1.1\textwidth}{!}
		{\includegraphics{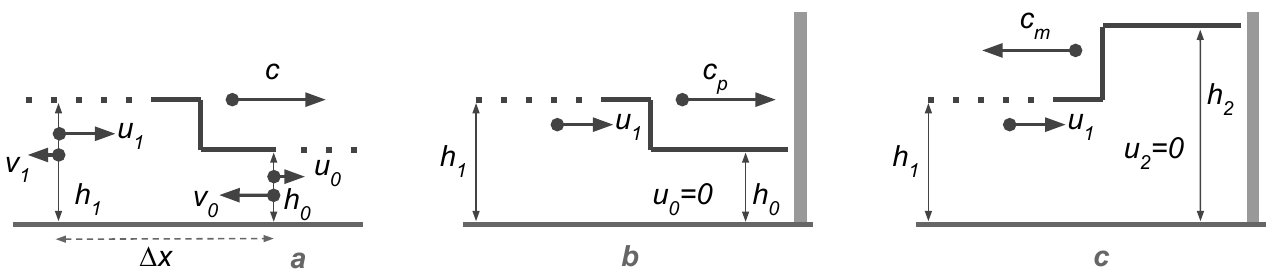}}
	\caption{Shock geometry diagram.}
	\label{basics}
\end{figure}
A one-dimensional (that is, uniform in the cross-flow direction) bore or shock -- a propagating discontinuity of flow conditions -- can be characterized by five parameters:  flow depth $h_0$ and velocity $u_0$ on one side of the shock, flow depth $h_1$ and velocity $u_1$ on the other side of the shock, and the shock propagation speed $c$ (Figure \ref{basics}, a). Hereafter, all velocities refer to a depth-averaged velocity component in the direction of propagation $x$, positive upriver; the cross-flow axis will be denoted $y$, and the vertical axis -- $z$. The five parameters satisfy two relations formulated in terms of flow velocities relative to the shock front $v_0=u_0-c$ and $v_1=u_1-c$:
\begin{eqnarray}
\label{sh1}
h_0 v_0&=&h_1 v_1=m, \\
\label{sh2}
h_0 v_0^2 - h_1 v_1^2&=&m(v_0-v_1)=g(h_1^2-h_0^2)/2 \pm \sigma \Delta x,
\end{eqnarray}
where $m$ is the volume flux across the shock front. 
These relations, known as the shock conditions, express mass and momentum balance in a system of reference moving with the shock  \cite[]{stoker, henderson}. The first condition equates the mass flows across the unit width of the shock. The second condition equates the rate of the momentum gain by a block of fluid of length $\Delta x$ and unit width enclosing the flow discontinuity, and hydrostatic thrust on the block's faces plus action of friction; $h_{0,1}$ and $v_{0,1}$ being the flow parameters on the block's faces; $\sigma$ being the bottom shear stress; fluid density is taken as unity.
In our case, depths $h_{0,1}$ will be provided by 1-min-sampled water level measurements. 
In a system of reference moving with the shock at, say, 6 $m/s$ celerity, this measurements would be taken at a distance $\Delta x=360$ m apart. Friction can still be neglected, if it is small compared to the hydrostatic trust, which sets a lower limit on the shock height. Let us estimate a shock height $\Delta h_{min}$, for which contributions of pressure and bottom friction in the momentum change are equal in magnitude.
First,
$g(h_1^2-h_0^2)/2=g \Delta h (h_0+h_1)/2 \approx g \cdot \Delta h_{min} \cdot h$. Equating that with the friction term and using Manning formulation $\sigma=gn^2u^2/h^{1/3}$ results in
$ \Delta h_{min} \approx n^2 u^2 \Delta x / h^{4/3}$, where $n$ is Manning roughness coefficient. In the Kitakami-type rivers, $n \approx 0.03$ $s \cdot m^{-1/3}$ \cite[]{stoker},  (\url{http://www.engineeringtoolbox.com/mannings-roughness-d_799.html}). The flow velocities, as estimated in section \ref{secflow}, are under 2 $m/s$ for the later waves, and the typical flow depth is $h=4$ m.  Substituting these values yields $\Delta h_{min}=0.2$ m. From now on, the last term in \eqref{sh2} will be omitted, but with an understanding, that the resulting equation is applicable to shocks with heights well above 0.2 m.

The second shock equation can now be simplified. 
Using  \eqref{sh1},
\begin{eqnarray}
m(v_1-v_0)=h_0v_0 \cdot v_1 - h_1v_1 \cdot v_0=v_0v_1(h_0-h_1)
\end{eqnarray}
so \eqref{sh2} takes another form:
\begin{eqnarray}
\label{sh22}
v_0v_1=g(h_0+h_1)/2 .
\end{eqnarray}
Relative velocities $v_0$ and $v_1$ have the same sign opposite to the sign of the shock speed $c$. The flow depth is always greater behind the shock, than in front on the shock \cite[]{stoker, henderson}.

In real rivers with irregular cross-sections, however, the axial flow momentum is not preserved in each $x-z$ vertical plane. To accommodate the cross-river momentum exchange in what is now an essentially two-dimensional flow even when the waveform appear uniform on $y$-axis, 
the shock conditions need to be integrated across the river, which yields \cite[]{chanson2012}:
\begin{eqnarray}
\label{sq3}
\bar{v}_0 A_0 &=& \bar{v}_1 A_1 \\
\beta (A_1 \bar{v}_1^2 -  A_0 \bar{v}_0^2)  &=& \frac{g}{2} \left( \bar{h}_0 A_0 - \bar{h}_1 A_1 \right)
\label{sq4}
\end{eqnarray}
where $A_0$ and $A_1$ are the flow cross-sectional areas in-front and behind the shock front, $\beta=\bar{v^2}/{\bar{v}}^2$ is the momentum correction coefficient  \cite[]{henderson} assumed for simplicity to be the same on both sides of the shock, and $\bar{h}_j$, $\bar{v}_j$,  and $\bar{v^2}_j$, $j=1,2$, are the cross-sectional averages. A cross-sectional average of a value $\phi$ on a corresponding side of the shock is computed as
\begin{eqnarray}
\label{sq5}
\bar{\phi}_j=\frac{1}{A_j} \int {\phi_j \cdot da_j} =\frac{1}{A_j} \int {\phi_j \cdot h_j dy}, \ \ \ A_j=\int {h_j dy}
\end{eqnarray}
where $h_j(y)=\eta_j-d(y)$, $d(y)$ is the bed elevation, and $\eta_j$ is the surface elevation above the reference level (TP, in our measurements) which is presumed constant across the river. Note, that $\bar{h}$ is not a common-sense average depth $A/b$, where $b$ is the river breadth.
Each integral is computed over a domain $h_j>0$, so the river width can vary with the river stage $\eta$. 

Then given the flow cross-sections before and after the shock and its celerity, the system \eqref{sq3}-\eqref{sq4} allows to find the discharge rate across the shock $M$, the average velocities $u_0$ and $u_1$, and the inflow rates $Q_0$ and $Q_1$ relative to the river banks, as
\begin{eqnarray}
\label{sq6}
M=\mp \sqrt{ gA_0A_1 \left( \bar{h}_1 A_1-\bar{h}_0 A_0  \right) / \left( 2 \beta (A_1-A_0) \right) } \\
\label{sq7}
u_j=c + M/A_j, \ \ \  Q_j= c A_j + M 
\end{eqnarray}
with `$-$' sign in \eqref{sq6} for a direct bore ($c>0$), and `$+$' for reflected ($c<0$).
Should the river breadth be the same on both sides of the shock, then
\begin{eqnarray}
 \bar{h}_1 A_1-\bar{h}_0 A_0&=&\int {(h_1^2-h_0^2)dy}=(\eta_1-\eta_0) \int {(h_1+h_0)dy}=  \nonumber \\
 &=&(\eta_1-\eta_0)(A_1+A_0),
 \label{sq10}
 \end{eqnarray}
since $\eta_j$ is constant across the river. Likewise, 
\begin{eqnarray}
A_1-A_0=\int {(h_1-h_0)dy}=(\eta_1-\eta_0) \cdot b
\label{sq11}
\end{eqnarray}
and \eqref{sq6} simplifies to
\begin{eqnarray}
\label{sq66}
M=\mp \sqrt{ gA_0A_1 \left( A_0+A_1  \right) / \left( 2 \beta b \right) }
\end{eqnarray}
Condition \eqref{sq4} can be transformed in a manner similar to deriving \eqref{sh22}. 
Should the river breadth do not change, the resulting equation becomes:
\begin{eqnarray}
\label{sq12}
\bar{v}_0 \bar{v}_1=g(A_0+A_1)/(2 \beta b) .
\end{eqnarray}

Equations \eqref{sq6}-\eqref{sq7} will be used to evaluate average velocities and discharge rates at Fukuchi and Iino stations in the 2011 tsunami event. Equations \eqref{sq3} and \eqref{sq12} will be exercised with reconstructing incident and reflected wave heights by the weir. 

\section{Flow velocity estimates}
\label{secflow}

\begin{figure}[ht]
	\includegraphics[width=\textwidth]{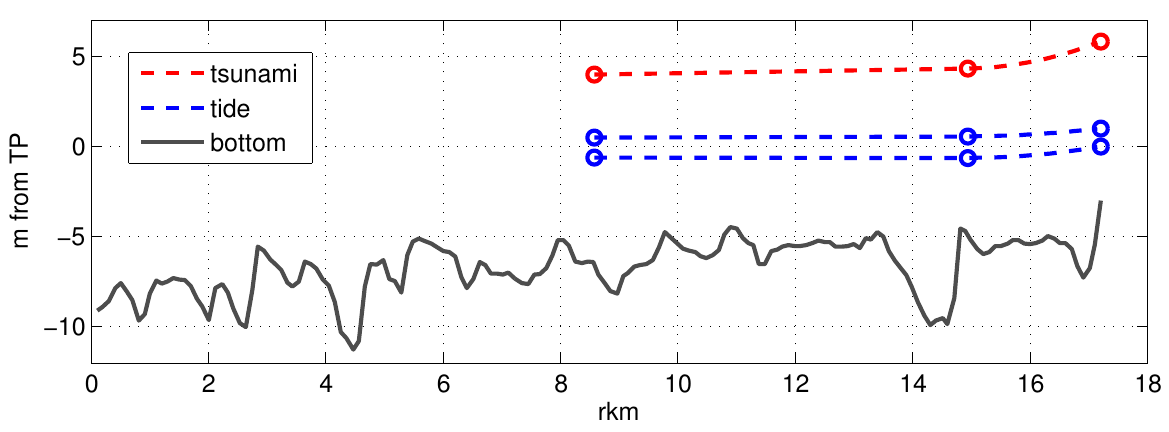} \\
	\includegraphics[width=\textwidth]{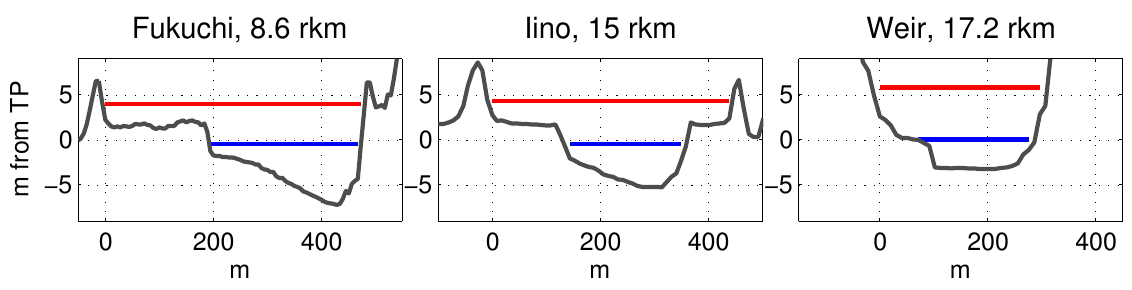} 
	\caption{Top: along-river profiles of the river bed (mouth-to-weir) and surface (Fukuchi-to-weir): gray - lowest bottom elevation; blue - high and low tidal levels on the day before the tsunami; red - maximal observed water level during the tsunami. Circles denote the measured water levels (left to right) at Fukuchi, Iino, and the weir. Bottom: cross-river profiles at the stations: gray - bed cross-section; blue - water level right before the tsunami arrival, red - maximal observed level.}
	\label{xsecs}
\end{figure}

\begin{figure}[ht]
	\includegraphics[width=\textwidth]{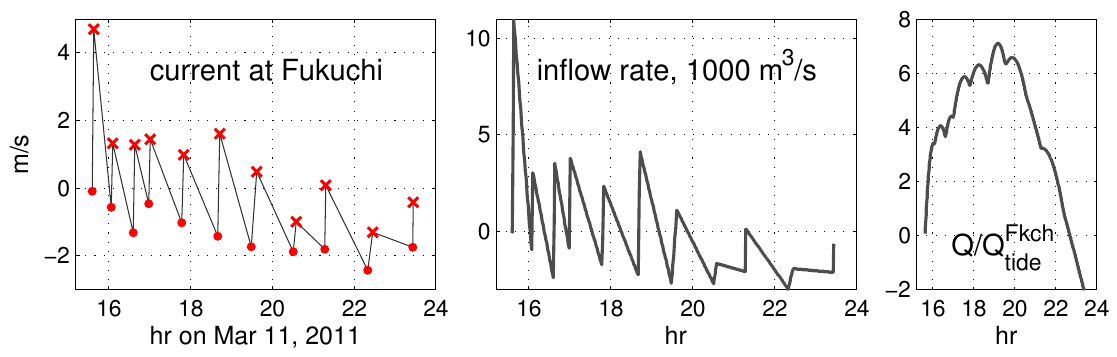} \\
	\includegraphics[width=\textwidth]{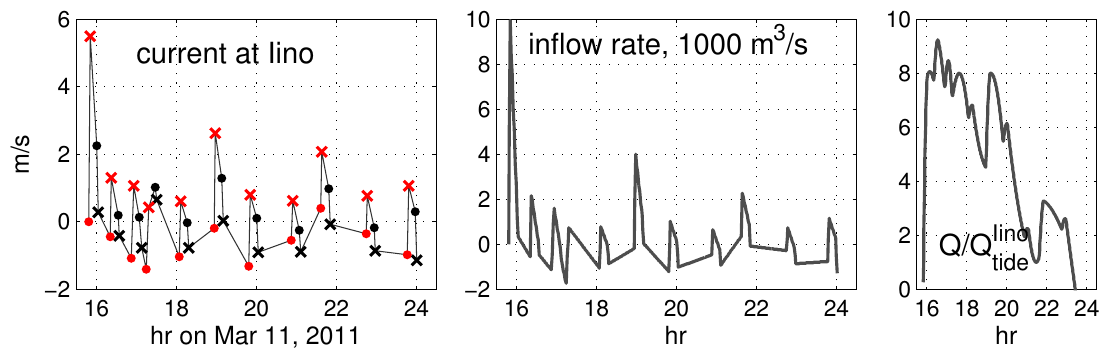} 
	\caption{Left: Estimated average flow velocity in front of the shock $\bar{u}_0$ (dots), and behind the shock $\bar{u}_1$ (crosses), for bores ascending the river (red) and reflected (black). Velocity is positive when directed upriver. Middle: inflow rate, in 1000 $m^3/s$. Right: inflow at the site since the tsunami arrival, as a fraction of a tidal prism on the previous day; tidal prisms at the sites are $Q^{Fkch}_{tide}=2.5 \cdot 10^6 \cdot m^3$, $Q^{Iino}_{tide}=0.54 \cdot 10^6 \cdot m^3$. Top row: Fukuchi; bottom row: Iino. }
	\label{flow}
\end{figure}

A tsunami record at a water level station readily provides two flow areas $A_0$ and $A_1$, given the bed profile across the river at the station. 
The Kitakami bathymetry data used in this work were collected by the Kitakamigawa-Karyu (Downstream region of Kitakami River) River Office, MLIT, and further processed in the National Institute for Land and Infrastructure Management (NILIM), Japan. The bed elevation in Kitakami was measured at the river's cross-sections at a 200-m interval along the river, and then interpolated onto a 10-m-resolution grid \cite[]{bathy}. Figure \ref{xsecs} shows the cross-river bed profiles next to the gauges, as well as the along-river bed and surface profiles.
The shock propagation speed has been obtained from the water level records, as described in section \ref{pspeed}. 
Flow estimates for Fukuchi were carried out with the designated eleven shock fronts, neglecting contribution from reflected waves. Coefficient $\beta$ in \eqref{sq6} was set to 1.5 for the first shock, and 1.25 for all later shocks. At Iino, both direct and reflected shocks were considered, with $\beta=1.05$ for direct shocks and $\beta=1.25$ for reflected. The greater value for the reflected shocks was set with an expectation that the flow in the reflected shocks, colliding with the tails of the direct ones, would be less uniform. 

The discharge rate obtained in front of the very first tsunami bore is that of the undisturbed background current comprised of riverine and tidal components. Let us estimate the background discharge using the continuity equation. For a channel of constant width, with a fixed inflow rate $q_0$ at a weir at $x=x_W$,
\begin{equation}
b \cdot \eta_t+q_x=0, \ \ \ q=bhu, \ \ \ q(x_W)=q_0
\label{cont}
\end{equation}
where subscript denotes partial derivatives. As seen in Figure \ref{zoom}, top, tidal signal is approximately the same from Fukushi to the weir, therefore $\eta_t$ is constant along  
the river. Then the solution to \eqref{cont} is:
\begin{equation}
 q(x,t)=q_0+\eta_t \cdot (x_W-x) \cdot \hat{b},
 \label{p}
 \end{equation}
where $\hat{b}$ is an average river breadth between the site and the weir. Approximating tide in the river with M2 tide 0.5 m in amplitude, we can estimate $\eta_t$ on the tsunami arrival 2 h after the low tide as $\eta_t(\tau)=0.5 \cdot \omega \cdot  \sin{\omega \tau}$, where $\omega=2 \pi /(12.42 \cdot 3600)$ $s^-1$, and $\tau=2 \cdot 3600$ s; then $\eta_t(\tau)=6 \cdot 10^{-5}$ $m/s$.
River breadth in-front of the first tsunami wave at Fukuchi was 280 m, and the distance to the weir was $x_W-x=8,630$ m; and at Iino, these numbers were 235 m and 2,260 m. The breadth at either location appears to represent an averaged breadth from this location to the weir. Then tidal inflow rate on the tsunami arrival evaluates to 140 $m^3/s$ at Fukuchi, and to 30 $m^3/s$ at Iino. Freshwater outflow from the weir before the tsunami varied around 100 $m^3/s$; that results in total inflow rates of about +40 $m^3/s$ at Fukuchi, and -70 $m^3/s$ at Iino. 
The inflow rates found with \eqref{sq6}-\eqref{sq7} are 
+8 $m^3/s$ at Iino and -96 $m^3/s$ at Fukuchi. However, using $\beta=1.0$ (instead of 1.05) at Iino, and $\beta=1.6$ (instead of 1.5) at Fukuchi changes these  rates to -167 $m^3/s$ at Iino and +162 $m^3/s$ at Fukuchi. 
That justifies the selection of $\beta$ within 0.05 for the first bore. For the later bores at Fukuchi, and the reflected bores at Iino, our choice of $\beta=1.25$ is merely our rough estimate for the flow ``non-uniformity", which we placed in-between a uniform flow and a flow under the first, inundating bore in Fukuchi.\\

The resulting inflow and average current estimates at Fukuchi and Iino, shown in Figure \ref{flow}, appear perfectly sound. In particular, the flow velocity estimates at Iino 
follow an expected sequence (Figure \ref{flow}, lower left): each direct bore sharply raises the flow velocity at the site (from red dots to red crests in the plot), the following reflected bore 12 min later finds the current reduced due to dissipation and adds negative increment to it (black dots to black crests), and so on, except for the 4-th reflected bore. The 4-th reflected bore had the smallest of all bores height 0.3 m, with the next smallest bore being 0.45 m high. Unrealistic results of the flow reconstruction around the 4-th bore might be due to missing the bottom friction in the momentum balance, as discussed in section \ref{algebra}.   

The corresponding inflow rates show that upstream Fukuchi, the river was mostly filling up for the first 3-3.5 hours, after which the drawdown prevailed, in spite of the later incoming waves. 
The plots on the right show the inflow into the river at each location starting from the tsunami arrival, computed by time-integrating the inflow rate at the site. Note that the total inflow into a river segment upstream each site should also include discharge from the weir. The inflow is expressed in tidal prisms\footnote{Tidal prism is a volume of water entering between mean low tide and mean high tide, or an average volume leaving at ebb tide.} upstream each location. The tidal prisms are estimated as follows.
Elevating the river stage through 1 m (the tidal range prior to the tsunami event) from Fukuchi to the weir takes $Q^{Fkch}_{tide}=8630 \cdot 290=2.5 \cdot 10^6 \ m^3$ of water to flow in by Fukuchi, where the average river width between Fukuchi and the weir is taken equal to the river width at Fukuchi at TP level. Likewise, elevating the river stage by 1 m upstream of Iino takes $Q^{Iino}_{tide}=2260 \cdot 240=0.54 \cdot 10^6 \  m^3$ of water to flow in by Iino. 
Elevating the river stage by 4-5 m over 1.5 river width (considering inundation) would accommodate 6-7.5 tidal prisms, so the resulting inflow estimates appear reasonable. 
However, the inflow estimates are still not reliable, because of sensitivity to even small errors which include/arise from a systematic error of replacing a shock speed $c$ at a site with an average speed between the sites, uncertainty of the coefficient $\beta$, and neglecting bottom friction. These errors, if small, introduce only a small bias to the current and inflow rate estimates, but accumulate during the integration. For the average current estimates from water level records, however, the method employed here could be fairly reliable.

\subsection{Bore or undular bore?}

\begin{figure}[ht]
	\resizebox{\textwidth}{!}
		{\includegraphics{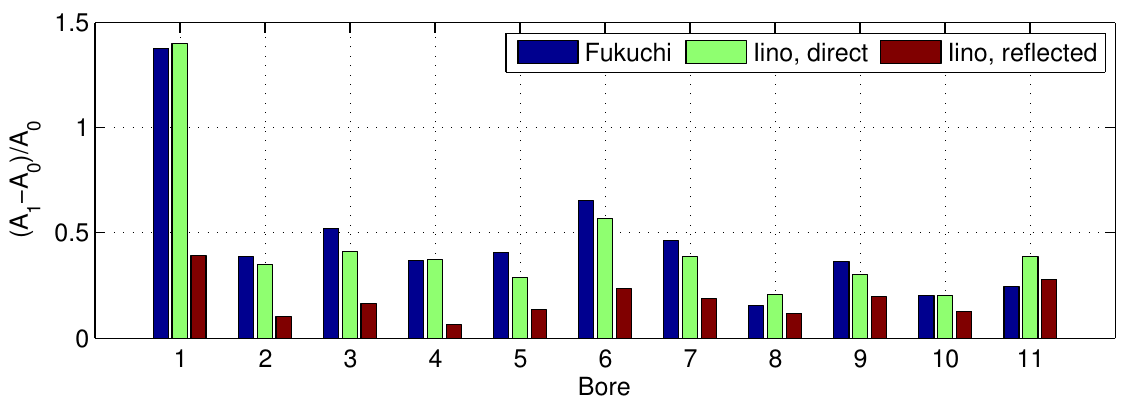}}
	\caption{Bore strengths $(A_1-A_0)/A_0$ for the eleven bores at subsequent locations.}
	\label{dratios}
\end{figure}

Tsunamis were sometimes observed to ascend in rivers as undular bores. Tsuji et al. (1991) and Yasuda (2010), respectively, show photographs of undular bores of the 1983 Japan Sea tsunami in a channel in the Noshiro Port and of the 2003 Tokachi-Oki tsunami in the Tokachi river. Describing their survey of the 2004 Indian Ocean tsunami in rivers in Sri Lanka, Tanaka et al. (2008) note that ``a witness at the 12 km point [up the Kalu river] testified that a big wave was immediately followed by 3 to 4 smaller waves traveling up the river, suggesting formation of soliton fission". 

Undular bores form when energy loses in the system are low  \citep{benjlight}, which is linked to low ratios of the conjugate bore depths (or, in our case, cross-sectional areas) $A_1/A_0$. Experimental study by Chanson (2010) found undular bore formation whenever a ratio of the conjugate depths was less than 1.7-2.1. However, these numbers, obtained in a flat wave flume, might not directly apply to rivers with irregular bed shapes. A bore propagating in a channel with a complex cross-section would disintegrate, if the axial flow momentum were preserved in each vertical plane - therefore, the momentum must be redistributed across the river. Cross-river momentum transfer is facilitated by the presence of turbulence on the bore front, which implies greater energy losses than the same bore might have in a flat channel. It is anticipated therefore, that the complexity of the river bed shape might act toward increasing dissipation and thus impeding formation of undulations.

The ratios  $(A_1-A_0)/A_0$, also called bore strength, for all bores at consequent locations in this study are plotted in Figure \ref{dratios}. The bores kept their strength relative to each other: the stronger bore at Fukuchi remained stronger at other check-points later. The ratio $A_1/A_0$ is below 1.7 for all bores except the first one. Should the undulations had occurred, their period would be well below the sampling interval of 1 min, therefore they might reveal themselves only as a large (up to 0.5 bore height) discrepancy between two neighboring measurements on a bore's top, in particular, as a one-node peak on the top of a bore. The latter is not observed. Still, 1-min sampling interval is too coarse to reliably confirm or rule out the undulations. 

\section{Wave heights near the weir}

Waves measured immediately downstream the weir are formed by the superposition of what was the direct wave and what will be the reflected wave. 
Stoker (1957) described a method of computing a relative height of a reflected shock $h_2/h_0$, given a relative height $h_1/h_0$ and flow velocity $u_1$ behind an incident shock, while $u_0=u_2=0$ (Figure \ref{basics}, b-c). Neglecting the riverine flow, as well as any subsequent background current by the weir before the next bore arrival, as well as escape of mass and momentum over the weir, our computations closely follow Stoker's method, with a slight modification due to a different set of unknowns, and due to using the cross-river averaged model rather than the original 1-D model. Namely, given a record by the weir (Figure \ref{weir}) and the bottom profile (Figure \ref{xsecs}, right panel), we know the flow areas before and after the reflection $A_0$ and $A_2$, while looking to estimate the heights of the incident and reflected bores $\eta_1$ and $\eta_2$, and the celerities of the direct and reflected shocks $c_p$ and $c_m$ (Figure \ref{basics}, b-c). Assuming the same river's breadth before and after each shock, we will use equations \eqref{sq3} and \eqref{sq12}.

Substituting $A_0=A_1 \cdot \bar{v}_1/\bar{v}_0$ into \eqref{sq12}, and substituting $\bar{v}_1=\bar{u}_1-c$, $\bar{v}_0=-c$ for a shock approaching a wall (since $u_0=0$), yields
\begin{eqnarray}
\label{sh3}
-c(\bar{u}_1-c)=\frac{gA_1}{2 \beta b} \left( 1-(u_1-c)/c \right)
\end{eqnarray}
Equation \eqref{sh3} can be re-written as a cubic equation for a relative shock speed $\zeta=c/\bar{u}_1$, function of a dimensionless combination $\alpha=gA_1/(\beta \cdot b \cdot \bar{u}_1^2)$:
\begin{eqnarray}
\label{sh4}
\zeta^3-\zeta^2-a\zeta+a/2=0
\end{eqnarray}
This equation has three distinct real roots, one positive, one negative, and a third root which has a value between the other two and which is not physically meaningful \cite[]{stoker}. An incident shock celerity $c_p$ therefore corresponds to the largest root $\zeta_p>0$ of \eqref{sh4}, and a reflected shock celerity $c_m$ -- to the smallest root $\zeta_m<0$. 
As follows from  \eqref{sq3}, flow areas relate to the shock celerity as:
\begin{eqnarray}
\label{sh5}
A_1/A_0  &=&  -c_p/(\bar{u}_1-c_p)=\zeta_p/(\zeta_p-1) =  f_1(\alpha) \\
\label{sh6}
A_2/A_1 &= &  (\bar{u}_1-c_m)/(-c_m)=(\zeta_m-1)/\zeta_m =  f_2(\alpha) \\
\label{sh7}
A_2/A_0  &= & f_1(\alpha) \cdot f_2(\alpha) =  f_3(\alpha)
\end{eqnarray}
Next, functions $\zeta_p(\alpha)$ and $\zeta_m(\alpha)$, $f_1(\alpha)$ and $f_3(\alpha)$ have been tabulated by solving \eqref{sh4} in a range of values $\alpha$ and then applying equations \eqref{sh5}-\eqref{sh7}. Now given $A_2$ and $A_0$, a corresponding parameter $\alpha$ can be found from \eqref{sh7}, and all other shock parameters can be found in a sequence:
\begin{eqnarray}
A_1=A_0 \cdot f_1(\alpha), \ \ \bar{u}_1=\sqrt{gA_1/(\beta b \alpha)}, \ \ c_p=\bar{u}_1 \cdot \zeta_p(\alpha), \ \ c_m=\bar{u}_1 \cdot \zeta_m(\alpha).
\label{sh8}
\end{eqnarray}
Function $A(\eta)=\int {(\eta-d(y))dy}$ has been tabulated, and used to find surface elevation $\eta_1$ given flow area $A_1$. 
As seen from this algorithm, particular values of the river breadth and parameter $\beta$ affects only the velocities, and do not affect the shock heigh estimates. 
The river breadth for each bore was computed as an average between its breadth at the bore's foot and that at its top, while $\beta$ was set to unity. 

\begin{figure}[ht]
	\includegraphics[width=\textwidth]{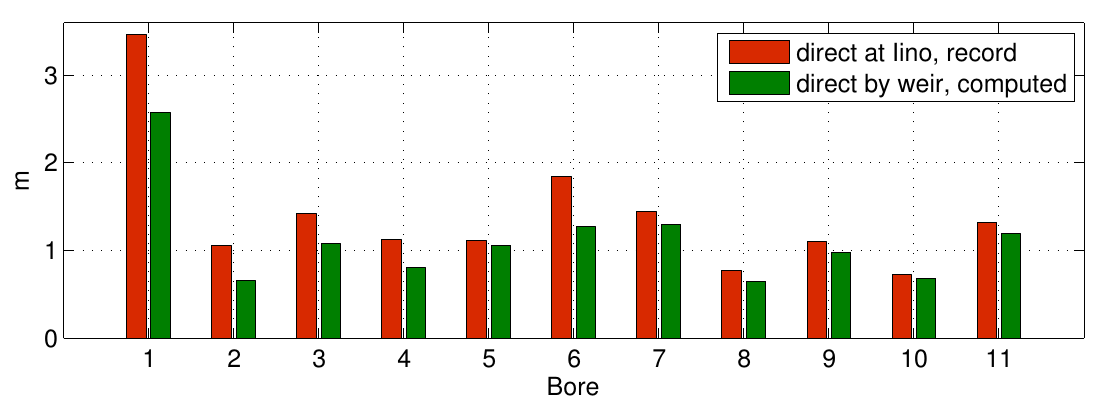} \\
	\includegraphics[width=\textwidth]{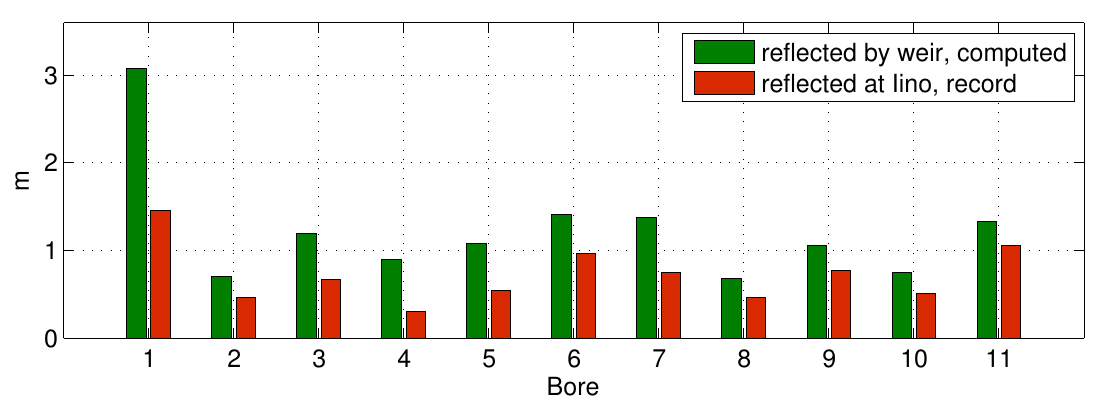} 
	\caption{Top: measured at Iino and estimated by the weir heights of the first eleven tsunami bores in the direct wave train. Bottom:  estimated by the weir and measured at Iino heights of the eleven bores in the reflected train. The bore height earlier in the travel is plotted to the left of its subsequent height. Red - measurements, green - calculations.}
	\label{bars}
\end{figure}
\begin{figure}[ht]
	\includegraphics[width=\textwidth]{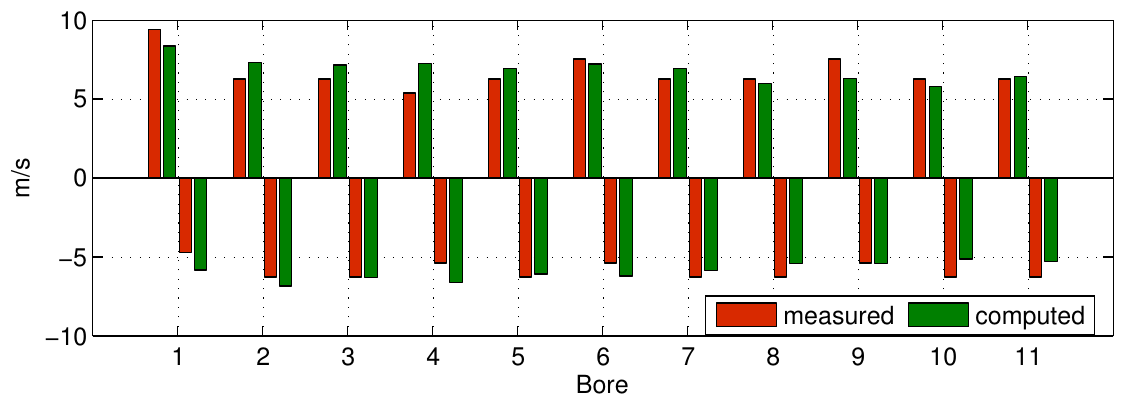} 
	\caption{Measured propagation speeds of the eleven bores on the way Iino-weir and weir-Iino, and estimated speeds of direct and reflected bores deduced from the water level record at the weir. Red - measurements, green - calculations.}
	\label{hvels}
\end{figure}

As seen in Figure \ref{basics}, c, the flow depth $h_2$ behind the reflected shock accommodates both the direct shock $\eta_p=h_1-h_0=\eta_1-\eta_0$ high and the reflected shock $\eta_m=h_2-h_1=\eta_2-\eta_1$ high. 
The reflected shock is always higher than the incident one; and it is higher, the higher the shock height relative to depth in front of it \cite[]{stoker}. 
A bar-plot in Figure \ref{bars} shows heights $\eta_p$ of each direct bore as it passes Iino (measured) and approaches the weir (estimated), 
and the heights of the reflected bores $\eta_m$ as they depart the weir (estimated) and come back to Iino (measured). The shocks dissipate and therefore diminish in height as they propagate. In this sense, the computed estimates of the wave heights near the weir appear consistent with the observations at Iino for all waves. The reflected waves seem to experience greater dissipation than the direct ones, as suggested both by the present wave height reconstruction, and by the fact that little of the reflected waves got registered at Fukuchi. 

Estimated individual bores' celerities, plotted in Figure \ref{hvels}, slightly deviate from the celerities computed with the observed arrival times in section \ref{pspeed}, which should be expected due to a number of adopted simplifications, including neglect of the background current near the weir. Effects of the background current on the average wave speed approximately cancel out for a back-and-forth passage. 
Adopting $c_p$ and $c_m$ as the estimates for the direct and reflected wave celerities between Iino and the weir, an average propagation speed on the way Iino-weir-Iino can be estimated as
\begin{eqnarray}
<c>=  2c_p|c_m|/(c_p+|c_m|)
\label{cave}
\end{eqnarray}
Average celerities of the eleven waves computed with \eqref{cave} vary between 5.4 $m/s$ and 7.1 $m/s$, with the ensemble average being 6.3 $m/s$, which matches the observed ensemble average speed on the way Iino-weir-Iino (see section \ref{pspeed}).

\section{Conclusions}

The 2011 Tohoku tsunami entered the Kitakami river and propagated there as a train of shock waves, recorded with 1-min interval at water level stations at Fukuchi, Iino, and the weir 17.2 rkm from the mouth, where the bulk of the wave was reflected back.  
The records showed that each bore kept its shape and identity as it traveled 10.9-rkm-path Fukuchi-Iino-weir-Iino. 
Shock handling based on the cross-river integrated classical shock conditions was applied to reconstruct the flow velocity time histories at the measurement sites, to estimate inflow into the river at each site, to evaluate the wave heights of approaching and reflected tsunami bores near the weir, and to estimate propagation speed of the individual bores. Two main objectives have been pursued.
The first objective was to test applicability of the SW (hydrostatic) approximation for describing tsunami bores with heights ranging from a few tens of cm to those exceeding the river depth, in a real river rather than in an idealized 1D channel. The second objective was to quantify different physical aspects of the tsunami intrusion. Our resources to reach both objectives were limited by the available field evidence, therefore our deductions about soundness of the theoretical estimates are often based on mutual consistency within the larger set of estimates or measurements, rather than on an immediate comparison with a direct measurement of the same value. 

We obtained very reasonable flow velocity reconstruction at Fukuchi and Iino. Flow reconstruction at Fukuchi showed, that amount of water which entered the river by Fukuchi is 7 times the tidal prism upstream Fukuchi. This water and the upstream discharge had been accumulating in the river's channel between Fukuchi and the weir for 3 hr after the tsunami arrival, before the drawdown began.
Given a water level record downstream the weir, we used the shock conditions to calculate the heights and celerities of the direct and reflected bores near the weir. The resulting height estimates by the weir are consistent with the measured direct  and reflected shock heights at Iino. The calculated average bore celerity in the path Iino-weir-Iino 6.3 $m/s$ exactly matched the measured one. 
Consistency of the obtained estimates suggests that the shallow-water theory can satisfactory describe the observed tsunami flow. Likewise, numerical simulations with a 2-D shallow-water model, performed in the Tohoku University, have demonstrated good agreement between the simulated water level variations and those obtained during a lab experiment imitating tsunami intrusion into a 1:330 scaled model of the lower 10-km long segment of the Kitakami river \cite[]{aoyama}.
At the same time, shock handling as above provides a simple and robust alternative to the full-event modeling for evaluating tsunami flow conditions, given the water level measurements along a river. The shock theory in application to tsunami bores in a river can be used in hazard mitigation, in particular, for the flow velocity estimates, which are the key factor in evaluating forces on bridges and other riverine infrastructure.
The accuracy of these estimates might be improved with more/better instrumentation targeting more frequent spacial coverage and more rapid sampling than even 1 min. 

\section{Acknowledgements}
The water level measurements were provided by the Kitakamigawa-Karyu (Downstream region of Kitakami River) River Office of the Tohoku Regional Bureau, Ministry of Land, Infrastructure and Transport of Japan.
The Kitakami River bathymetry was developed in the National Institute for Land and Infrastructure Management, Japan, for their collaborative research with the Tohoku University, using bed elevation data acquired by the Kitakamigawa-Karyu River Office. Authors thank two reviewers for their careful reading and helpful suggestions on improving the paper's presentation.

\section{Appendix. Separating direct and reflected wave trains in the Iino record}

 \begin{figure}[ht]
	\resizebox{\textwidth}{!}
		{\includegraphics{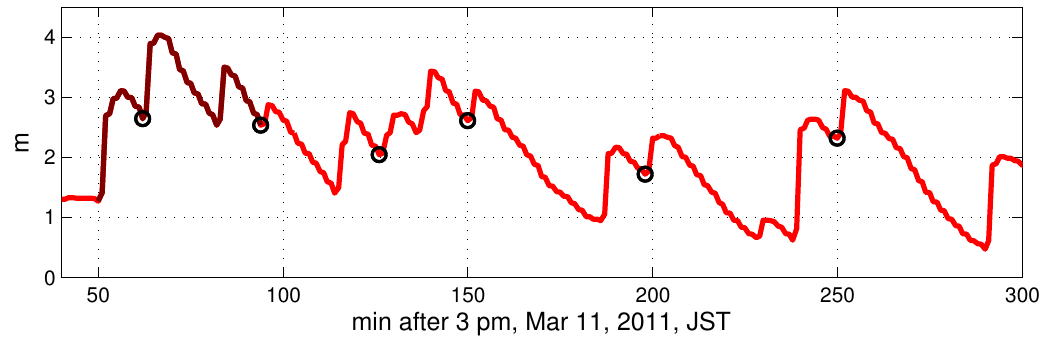}}
	\caption{Tsunami record at Iino. Black circles mark arrivals of reflected waves. Dark red: a record segment starting at the foot of the first direct wave and ending at the foot of the second reflected wave.}
	\label{iinoexpl}
\end{figure}

Consider the Iino record $y(t)$ as a sequence of superimposed incident and reflected waves, with the latter thought of as delayed and scaled copies of the incident waves. The delay time $\tau$ is found to be the same throughout the record and equal to 12 min, whereas the scaling factor $\alpha$ can vary among individual reflected pulses. Then
\begin{eqnarray}
\label{w1}
y(t)=s(t)+\alpha(t) \cdot s(t-\tau), \ \ \  \alpha(t)= \begin{cases}
0, & t<t_1 \\
\alpha_i, & t_i \le t<t_{i+1} \\
\end{cases}
\end{eqnarray}
where $t_i$ is an arrival time of an $i$-th reflected wave, and $s(t)$ is the incident wave train. The arrivals at times $t_i$ are marked with circles in Figure \ref{iinoexpl}.
For any specific segment of the record, \eqref{w1} represents a well-defined linear system, which can be written as
\begin{eqnarray}
\label{w2}
y&=&M \cdot s,
\end{eqnarray}
with matrix $M$ having all zeros, except ones on the main diagonal and $\alpha$ on a diagonal located $\tau$ rows below the main one. 
For instance, 
consider a segment of the record starting at the foot of the first direct wave and ending at the foot of the second reflected wave (dark red in Figure \ref{iinoexpl}). For this segment,
\begin{equation}
M =  \left[
\begin{array}{ccccc}
1 & \ldots & 0 & \ldots & 0 \\
\vdots & \ddots & \vdots & & \vdots  \\
\alpha_1 & \ldots &1  & \ldots & 0\\
\vdots & \ddots & \vdots &  \ddots & \vdots \\
0 & \ldots &\alpha_1 & \ldots &1\\
\end{array}
\right ]
\label{w3}
\end{equation}
Hence the corresponding segment of the incident wave can be found as $s=M^{-1} \cdot y$, while the reflected wave can be found as $r=y-s$. Since it is expected that the bores keep their shape, coefficient $\alpha_1$ is selected to maximize visual correlation between $s(t)$, $t_1 \le t <t_2$, and a corresponding segment of the record by the weir.

This procedure can be continued by extending the record's segment until the next reflected wave arrival. 
On each extension to $t_i \le t<t_{i+1}$, only the next  value $\alpha_i$ needs to be determined, with $\alpha_1, \dots, \alpha_{i-1}$ being already found. Proceeding in this manner allows to grow the incident and reflected wave trains by a wave at a time, extending the computed signals in $t_{i+1}-t_i$ increments. 


\end{document}